\documentclass[graybox,nosecnum,natbib]{svmult}

\bibpunct{(}{)}{;}{a}{}{,} 

\pdfoutput=1   

\usepackage{mathptmx}       
\usepackage{helvet}         
\usepackage{courier}        
\usepackage{type1cm}        

\usepackage{makeidx}         
\usepackage{graphicx}        
\usepackage{multicol}        
\usepackage[bottom]{footmisc}
\usepackage[normalem]{ulem}	
\usepackage{hyperref}  
\usepackage{amssymb}

\usepackage{soul}   

\newcommand*\aap{A\&A}

\newcommand*\aj{AJ}

\newcommand*\apj{ApJ}
\newcommand*\apjl{ApJ}

\newcommand*\apjs{ApJS}

\newcommand*\apss{Ap\&SS}

\newcommand*\icarus{Icarus}

\newcommand*\mnras{MNRAS}

\newcommand*\nar{New A Rev}
\newcommand*\nat{Nature}

\newcommand*\pasp{PASP}

\newcommand{\hbindex}[1]{{#1}}  

\makeindex             

\begin{document}

\title*{Circumstellar discs: What will be next?}
\author{Quentin Kral, Cathie Clarke  and Mark Wyatt}
\institute{Institute of Astronomy, University of Cambridge, Madingley Road, Cambridge CB3 0HA, UK, \email{qkral@ast.cam.ac.uk,cclarke@ast.cam.ac.uk}}
%
%
\maketitle

\abstract{This prospective chapter gives our view on the evolution of the study of circumstellar discs within the next 20 years from both observational and theoretical sides. We first present the expected improvements
in our knowledge of protoplanetary discs as for their masses, sizes, chemistry, the presence of planets as well as the evolutionary processes shaping these discs. We then explore the older debris disc stage and explain what will
be learnt concerning their birth, the intrinsic links between these discs and planets, the hot dust and the gas detected around main sequence stars as well as discs around white dwarfs.}


\section{Protoplanetary discs}

\begin{svgraybox}
\hbindex{Protoplanetary discs} are the discs of gas and dust that surround  a
significant fraction of stars  with ages  less than a few - 10 Myr.  The
best studied examples are located in nearby star forming clouds (typically
around 150 pc from the Earth) but evidence of discs is found in more distant
clusters and even in star forming regions in  the Magellanic Clouds.
Protoplanetary discs  are often termed `primordial discs' in order to
indicate that they are composed of material from the local interstellar medium which has collapsed into  a centrifugally supported disc around the
young star. Their typical gas and dust inventories (tens of Jupiter masses
and tens to hundreds of Earth masses respectively) imply that at this stage
they have sufficient material to form (exo-)planetary systems. Nevertheless,
it is clear that not all the material in protoplanetary discs is
destined to turn into planets. There is good evidence that there are significant
accretion flows from the discs onto their central stars while winds, either
magnetohydrodynamical or thermal, are likely to play a role in dispersing
such discs after a few Myr.
\end{svgraybox}

\subsection{The properties of protoplanetary discs}

\subsubsection{Disc masses}
 Currently there is considerable uncertainty about the robustness of the
techniques used to assess disc {\it gas} mass. In default of better indicators, the
traditional approach has been to assess dust mass from the mm continuum flux and
convert to a total gas mass by
assuming  a dust to gas ratio of 1:100 as in the interstellar medium.
This is a questionable assumption since there are a variety of processes
that can drive differential evolution of dust and gas within discs. Direct
gas measurements are however hard:
since the infrared lines of molecular hydrogen (the dominant gas species)
are weak and do not trace the bulk of the gas mass \citep  [e.g.][] {2013ApJ...779..178P},
proxy molecules, particularly
CO, are used instead.
Although the most abundant \hbindex{isotopologue} of CO ($^{12}$CO) cannot be
used as a mass tracer because it is usually optically thick, the
high sensitivity of
\hbindex{ALMA} permits  measurements of
the rarer isotopologues,  $^{13}$CO and C$^{18}$O \citep {2014ApJ...788...59W} and these
are converted into total gas mass assuming an abundance  equal to that in the
dense interstellar medium \citep{1994ApJ...428L..69L}. CO based total gas masses are however
much lower than those
obtained from dust measurements and,
when combined with measurements of accretion onto the central star,  can
imply disc lifetimes much less than the system age  \citep {2016A&A...591L...3M}.
While the problem can be partially mitigated by  careful treatment of photodissociation
and freeze-out of CO \citep [e.g.] [] {2014A&A...572A..96M}, the low
values of CO based gas mass estimates have been challenged by  HD 1-0 line measurements
from  \hbindex{Herschel}
\citep{2013Natur.493..644B} which  imply gas masses that are a factor $3$ to $100$ higher than
those based on CO. Current modelling \citep{2017trapman} suggests that carbon depletion
is the most likely source of this discrepancy.

\begin{figure}
\centering
\includegraphics[scale=0.6]{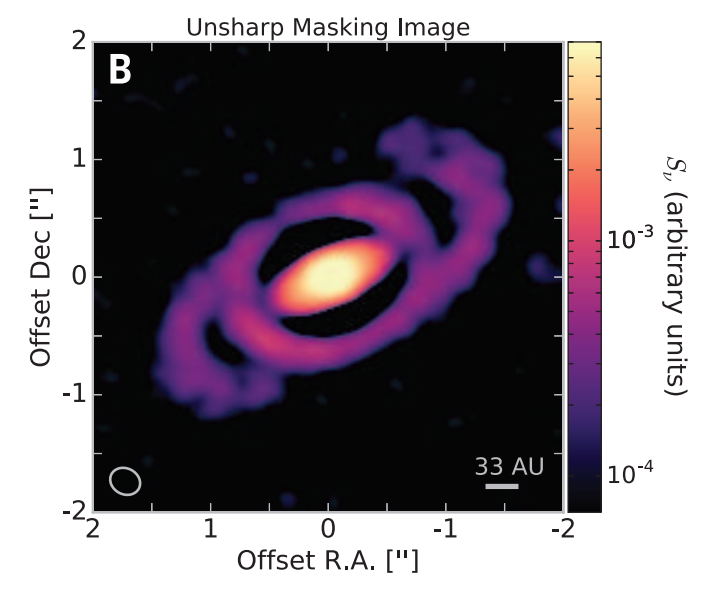}
\caption{An example of spiral structure on a scale of 100-200au in the young star \hbindex{Elias
227} obtained by applying unsharp masking to a high resolution ALMA continuum image \citep{2016Sci...353.1519P}.}
\label{figgi}
\end{figure}

 This uncertainty about  disc gas masses implies uncertainties about
the environment of forming protoplanets and
even  the planet formation  mechanism. Disc mass is a crucial
discriminant between the two main competing models for gas giant planet
formation (i.e. core accretion versus gravitational instability; 
\citealt{2007prpl.conf..591L, 2007prpl.conf..607D}), since the
latter requires disc masses that are in the region of $10 \%$ of the central
stellar mass. High resolution imaging (e.g. in
the submm continuum with ALMA) can however
provide an alternative tool for assessing whether the conditions for
\hbindex{gravitational instabilty} are in fact being met (see \citealt{2016Natur.538..483T} for evidence
of a disc undergoing gravitational fragmentation,
albeit on a stellar rather than planetary scale
and \citealt{2016Sci...353.1519P}  (Fig.~\ref{figgi}) for disc \hbindex{spiral} structure  which may
be interpreted as evidence for gravitational instability in the disc).

 Ultimately, the issue of disc gas mass will not be settled definitively until HD 1-0
measurements become available for a broader range of systems and even then their
interpretation will need a sophisticated modelling effort combined with information
from high resolution mapping in other tracers such as CO. From 2018, further
HD measurements will become available using  the \hbindex{HIRMES} instrument
aboard \hbindex{SOFIA} \citep{2017trapman}, while there are more distant possibilities of achieving still higher
sensitivity via the proposed \hbindex{SAFARI} instrument on \hbindex{SPICA} \citep{2014SPIE.9143..1KR}. 

\subsubsection{Disc radii} 
 In recent years, it has become possible to measure disc outer radii
in both gas (CO lines) and dust (generally mm continuum). The
increased spatial resolution and sensitivity of mm arrays has allowed
surveys to sample the {\it diversity} of protoplanetary disc sizes: while
the largest discs extend over many hundreds of au  
\citep{2011A&A...529A.105G, 2013A&A...549A..92G} 
there are other systems for which the spectral energy distribution
as measured by Herschel implies  dust only within
an au of the star \citep {2014A&A...570A..29B}.
There is also a general trend for discs to
be more compact  in dust  than in gas 
\citep{2009A&A...501..269P,2014A&A...564A..95P},  a result that can be readily
understood in terms of inward radial migration of dust \citep{2014ApJ...780..153B}.
Indeed it has recently become apparent \citep[e.g.] [] {2014A&A...564A..95P}
that discs that are faint in the mm continuum
(which had previously been interpreted as having low gas mass) may instead be
merely compact (and thus optically thick in the dust continuum).
 
Currently there has been little attempt to fit the growing dataset of  disc outer radii into
a protoplanetary disc evolutionary scenario.  Outer disc radii are important to planet
formation for a number of reasons. First of all, for  typical surface density
profiles  \citep {2005ApJ...631.1134A}, the bulk
of disc  mass is contained at large radius, even if planets only form in the
inner regions of the disc,  their natal environments are probably re-supplied by flows
of gas  and solids from large radii.  Outer disc radii  also indicate  the processes
shaping disc evolution, which are important for the migration of
protoplanets in the disc. Protoplanetary discs are often modelled as
conventional viscous accretion discs in which radial inflow in the disc is driven
by redistribution of angular momentum to large disc radii resulting in  an {\it increase}
in outer disc radius with time \citep[e.g.][] {1998ApJ...495..385H}. On the other hand, recent 
modelling  suggests that the dominant evolutionary driver may instead
be large scale\hbindex{ magnetohydrodynamical winds} \citep  {2013ApJ...772...96B, 2013ApJ...775...73S}  which, in  removing angular momentum
from the disc, may instead cause the secular {\it decrease}  of disc radius with time.
A further complication is that even the rather mild ultraviolet radiation fields
in star forming regions are likely to drive significant  winds
from the outer regions of discs by \hbindex{photoevaporation} \citep[e.g.][]  {2016MNRAS.457.3593F}  which should cause
shrinkage of disc gas at late times \citep{2007MNRAS.376.1350C,2013ApJ...774....9A}.

\subsubsection{Grain growth}
An area of recent progress has been the quantification of  {\it \hbindex{grain growth}}
in protoplanetary discs. In the core accretion scenario it is a fundamental
tenet that grains grow from the sub-micron scales of dust in  the
interstellar medium to scales where they ultimately coalesce into planetary cores.
While it is still unclear whether this involves direct growth to \hbindex{planetesimal} 
(km scale) bodies or instead the accretion of
(roughly cm scale) `\hbindex{pebbles}'  \citep{2012A&A...544A..32L}, both variants imply
that discs should evidence grain growth.  Such evidence has been widely available
through compilation of submm spectral indices which are sensitive to the maximum
grain size in the dust population: if dust grows such that the size 
scales dominating the cross section are larger than
the wavelength of observation, the opacity becomes `grey' and thus
the spectral slope of optically thin emission simply follows the Rayleigh Jeans
dependence on wavelength.  Steeper spectral indices (more flux at shorter wavelengths)
however imply  frequency dependent opacity and hence more   compact grains.
While indications of grey opacity (and hence grain growth) in unresolved
observations of protoplanetary discs have been available for over a decade
\citep {2010A&A...521A..66R,2014prpl.conf..339T},
it is only with the superior resolution of ALMA that it has become
possible to conduct this experiment as a function of radius within individual discs
\citep{2016A&A...588A..53T}. Pilot studies provide strong evidence that the maximum
grain size {\it declines} with increasing radius. It is currently unclear whether
this indicates more efficient growth of grains {\it in situ} in the inner regions
of discs or whether the larger grains in the outer disc have already become
decoupled from the more tenuous gas at large radius and undergone rapid
inward radial migration \citep[e.g.] []{2014ApJ...780..153B}. 
Whatever the evolutionary
interpretation of this data, it is evident that ALMA has the capability to
address this issue in much larger samples in the coming years and that this
will provide good constraints on the solid material available to planet
formation at various disc radii. It is however worth noting that this method
cannot be applied in the innermost disc - less than about 10 au -  due both
to finite resolution and the impossibility of applying this technique
in regions of the disc that are optically thick in the
mm continuum. Finally, although ALMA promises a spectacular increase in our
understanding of grain growth to mm scales, it will be necessary to go to longer
wavelengths in order to constrain grain growth to larger ($>$ cm) scales: ngVLA and
the higher frequency bands of SKA offer the prospect of probing such large grains
in the coming decade. 

\subsubsection{Disc chemistry}

We now turn to the intense efforts in recent years to characterise radial
variations in the
{\it chemistry}  of protoplanetary discs. Here again, increases in resolution and
sensitivity will facilitate the  assessment of  chemical conditions
as a function of radius (rather than simply obtaining disc averaged quantities that
are difficult to interpret). An area that is currently in its infancy, due to
the difficulties of removing contaminating envelope emission, is the chemical
characterisation of the youngest, most deeply embedded discs (i.e. Class 0/I sources:
see below). Here, new theoretical studies are charting the chemical processing
of cloud abundance patterns that are to be expected during the early phases
of disc assembly \citep{2015MNRAS.451.3836D}, as well as possible chemical signatures of
shock heating in self-gravitating discs \citep{2011MNRAS.417.2950I}. Such studies can be confronted with
ALMA data \citep{2013MNRAS.433.2064D} and with compositional information on cometary ices in the solar system \citep[e.g.][]{2015A&A...583A...1L}
and will help clarify  the initial chemical conditions  during the
planet formation era.

Considering the large scale distributions of 
carbon and oxygen in discs, the main reservoirs of these elements in discs are in the form of
CO, CO$_2$ and H$_2$O, with  minority components in  carbide and silicate
grains. These species transition between their solid and gaseous phases at their respective
\hbindex{snow-lines} which are located, for typical disc model parameters in solar type stars
at $\sim 30$au, $5$au and $2$au. There has been considerable interest
in locating  CO snow-lines in protoplanetary discs (as marked by the steep
decline in C$^{18}$O emission and the associated  rise in species such as DCO$^+$ and
N$_2$H$^+$ \citep{2011ApJ...740...84Q,2013A&A...557A.132M,2013Sci...341..630Q,2015ApJ...813..128Q},
partly in order to obtain an anchor point for
modelling  disc temperature structure, assuming  that the CO snow-line should correspond to
a local disc mid-plane
temperature of around $20$K. This exercise is however somewhat complicated
by the fact that the sublimation temperature of CO can vary according to whether it is in the form of pure CO ice or mixed CO-water ice
\citep {2004MNRAS.354.1133C}.  
There is nevertheless the prospect that the location of
CO snow-lines in large samples of protostellar discs will provide one of the most direct
insights into the thermal evolution of the mid-plane regions of protoplanetary
discs \citep{2017MNRAS.tmp.115P}.

 Apart from their interest as disc thermometers, CO snow-lines mark the region where most
of the carbon in the disc makes the transition between solid phase (CO ice) and
gaseous CO. 
Inward of the  snow-line,
around $50 \%$ of the oxygen in the disc enters the gas phase as gaseous CO, while,
outward of the water snow-line on a scale of  $\sim 1$au, the remainder resides
in solid form  as water ice.

 This division of carbon and oxygen between solid and gaseous phases at various radii
provides the chemical backdrop for planet formation and may in principle allow
determinations of elemental abundances in hot Jupiters to constrain planet formation
models. The mapping from formation scenario to chemical signature is however complex
and somewhat degenerate. While the division of carbon and oxygen into gaseous and solid
forms in various thermal regimes is reasonably well understood, planet formation
involves a complex interplay between the accretion of gaseous and solid phases along
with planetary migration. In addition,  elemental abundances are almost certainly
functions of
radius on account of the fact that  loosely coupled icy dust grains drift inwards relative
to the gas;  at the same time this drift  can be impeded by
grain accumulation
in pressure maxima and near ice-lines \citep{2008ApJ...685..584I}.
Analysis of elemental abundances of material accreting onto the star \citep
{2005ApJ...627L.149D,2013ApJS..207....1A,2015A&A...582L..10K}
 provides some evidence that objects with structures
where dust accumulates do show evidence for chemical filtration (i.e. a deficit
of elements that are retained in the disc in the form of grains or icy grain mantles).
The relationship between chemical fractionation, resolvable dust structures
in discs and the chemical composition of planets is clearly an area that requires
more thorough exploration \citep{2014ApJ...794L..12M,1611.03083,2017MNRAS.460L..10B}.

Overall,  chemical studies of protoplanetary discs will derive great benefit 
from the fact that the high resolution spectroscopy provided by ALMA in the submm and 
\hbindex{METIS} in the mid-infrared are well matched in spatial resolution (around 50 mas).
This will
permit a spatial dissection of disc
chemistry on scales of a few au in the closest star forming regions and will thus
directly access chemistry in the planet forming regions of discs.
  
\subsection{Evidence for planets in protoplanetary discs}
There are currently relatively few examples of planets discovered in protoplanetary
discs by the methods that are traditional for older planetary systems: evidently
the large radial optical depth of protoplanetary discs rules out
transit methods during the disc bearing phase. Radial velocity detections are
impeded by false signals owing to the active surface features on young stars, whereas
imaging studies need to contend with the possibility of false positives generated by
scattering off clumpy disc features. The former problem can to some extent be
mitigated by near-infrared spectroscopic monitoring, allowing starspot signals to
be distinguished from planets:  this has led to the first claimed detection
of a
radial velocity planet in a  disc bearing
star \citep[\hbindex{CI Tau},][]{2016ApJ...826..206J}. At an age of only around a Myr this provides
important evidence that at least some gas giant planets  can form and migrate to
small orbital radii at a very early stage of protoplanetary disc evolution. It is
currently unclear whether the moderate eccentricity of this planet ($\sim 0.3$)
is to be understood in terms of pumping by the massive disc or whether it requires
the presence of sibling planets \citep{2017MNRAS.464L.114R}.
There have also been
several claims of point source detections in protoplanetary discs which may
be interpreted as planets 
\citep[e.g.][] {2011A&A...528L...7H,2012ApJ...745....5K,2013ApJ...766L...1Q,2015Natur.527..342S,2015ApJ...808L..41T} 
though none have been confirmed thus far.

 The majority of information about possible planetary systems in disc bearing
stars derives from the effect that such planets have upon the disc. Occasionally
this is in the form of a stable periodic  feature in disc spectroscopy as in the
candidate hot Jupiter embedded in the vigorously
accreting young stars \hbindex{FU Orionis} \citep{2012MNRAS.426.3315P}. The majority of evidence
is however derived from disc structures imaged  either in scattered light (where it
represents a perturbation of  small grains at several scale heights above the disc)
or else in submm continuum (where it instead represents structure in large grains
close to the disc mid-plane).

 Many but not all of the growing compendium of imaged disc structures have been found
in systems that had been previously classified as `\hbindex{transition discs}' on account
of
their spectral energy distributions: these  evidenced missing emission
over a limited range of wavelengths, which suggested  the presence of annuli or holes
devoid of dust. For example \hbindex{TW Hydra} (see Fig.~\ref{fig:giant}), a system identified as a transition disc
from its spectral energy distribution 
\citep{2002ApJ...568.1008C}
has recently been shown to exhibit a wealth of annular structure
in near-infrared scattered light and submm imaging
\citep{2016arXiv161008939V,2016ApJ...820L..40A}. 
The term `transition disc' reflected  the belief that such discs belong to the
evolutionary phase where they make the transition between optically thick
protoplanetary disc status to debris disc phase (see next section).  This is now   thought 
to be not necessarily the case: many so-called transition discs are vigorously
accreting and have very high mm fluxes 
\citep {2012MNRAS.426L..96O, 2015MNRAS.450.3559N}, suggesting that
they instead belong to an early evolutionary stage. Modelling of planet carved
structures in discs initially involved only
 {\it gas} \citep[e.g.][]{1979MNRAS.186..799L}, where it was found that
significant perturbation in the gas
required rather massive planets \citep[around a  Jupiter mass or above,][]{2006Icar..181..587C}. 
It is now well known
\citep {2004A&A...425L...9P,2006MNRAS.373.1619R,2012ApJ...755....6Z,2014ApJ...789...59O,2014ApJ...785..122Z,2015ApJ...809...93D, 2015A&A...584A.110P} 
that observable structures in {\it dust} can be produced by
much lower mass planets which hardly perturb the gas. For planets more
massive than $20$ Earth masses, drag coupled dust is trapped
in the pressure maximum in the disc just outward of the planet and this trapping
is predicted to lead  to a {\it hole} in the mm dust distribution. Lower mass planets produce an
annular dust feature outside the planet's orbital radius but no interior hole.
Simulations \citep{2016MNRAS.459.2790R}  suggest that dusty structures in discs produced
by low mass planets should be detectable both in the submm continnum with ALMA and
in the infrared using either   existing  instruments
telescopes such as \hbindex{SPHERE} on the \hbindex{VLT} or \hbindex{GPI} on \hbindex{Gemini} or else in the thermal
mid-infrared  using future intruments such as METIS on the \hbindex{E-ELT}. The limits of detectability
(around $15$ Earth masses) are expected to be achieved with ALMA.

Apart from the annular structures discussed above, several transition discs  exhibit
spiral structure  in near-infrared scattered light imaging 
\citep{2012ApJ...748L..22M,2013A&A...560A.105G,2013Sci...340.1199V,
2013ApJ...762...48G, 2014ApJ...785L..12C,2015A&A...578L...6B,2015ApJ...813L...2W, 2016A&A...588A...8G, 2016A&A...595A.113S}. Although such spirals have been attributed to the presence of planets,
the interpretation is not straightforward \citep{2015MNRAS.451.1147J}  since the amplitude of surface density variations
produced by planets (typically several tens of per cent at most) is much less than the amplitude
of variation observed (a factor three or more). Planets may be able to
produce the  larger amplitude spirals seen
in scattered light through perturbation of the disc vertical scale height,
but in this case  the amplitude of spiral structure in the submm is predicted to be
much smaller. 

 It can be expected that the next decade will see a concerted attempt to compare
spirals seen in scattered light and mm continuum and to use these to try and constrain
the presence of planets in the disc. The largely null results that have emerged from
planet searches in  somewhat older (10-300 Myr) discless stars 
\citep{2013ApJ...777..160B, 2014ApJ...786....1B}
at large orbital radii ($> 50$ au) suggest that much of the structure is likely
to be on small scales. Here the  high resolution afforded
by ELT class telescopes in conjunction with ALMA will be essential. 

\subsection{Evolutionary processes  in protoplanetary discs}

 Young low mass stars have been traditionally classified according to
the slope of the infrared spectral energy distribution 
\citep{1984ApJ...287..610L}.  It is
now widely believed that the resulting classes, which have increasingly
steep spectral energy distributions (i.e. less flux at longer wavelengths)
represent an evolutionary sequence between objects that are strongly
disc/envelope  dominated (Class O/I) to those that are - at least approximately -
disc-less (Class III). The majority of disc bearing objects discussed
so far, and indeed the bulk of the disc bearing population, belong to
Class II, a stage where the star is clearly visible in the spectrum in
the optical but where there is also clear evidence of emission at
longer wavelengths, largely interpreted as stellar radiation reprocessed
by disc dust. Despite the uncertainties in disc masses
described above, it is very unlikely that the majority  of Class II discs
are sufficiently massive for the disc's {\it self-} gravity to be
important compared with that provided by the central star. At the younger
(Class 0/I) stage, this is not necessarily the case. Models for the collapse of protostellar cores \citep{2013MNRAS.433.3256V} 
predict
that discs should pass through an early self-gravitating phase which is
hard to access observationally, both on account of its relative brevity ($\sim
10^5$ years) and the fact that it coincides with a phase when protostellar
systems are deeply embedded in dust.  It is only extremely recently that
ALMA has started to reveal examples of objects that are probably in this stage
and which exhibit both massive spiral features and, in some cases, the formation
of fragments \citep{2016Natur.538..483T,2016Sci...353.1519P} . Such observations provide
vindication for the large body of theoretical and numerical
work which predicts that large  amplitude spiral structures (and hence the possibility
of disc fragmentation) should be restricted to the outer regions of
young protoplanetary discs ($> 50$ au) where the
ratio of the
cooling time to the dynamical time is short 
 \citep{2001ApJ...553..174G, 2005ApJ...621L..69R,2009ApJ...704..281R, 2009MNRAS.396.1066}: see however
\citet{2011MNRAS.411L...1M,2012MNRAS.427.2022M, 2011MNRAS.416L..65P,2014MNRAS.438.1593R,2015MNRAS.451.3987Y}  for a discussion
of the challenging numerical issues involved in modelling protostellar discs
during the self-gravitating phase.
 
 In the next decade it can be expected that ALMA will improve our understanding
of the earliest phases of disc evolution considerably. Although this phase  is
brief, it is  potentially long enough to allow the formation of planets by
gravitational instability in the outer disc \citep{2007prpl.conf..607D}.  Spiral modes in the disc can
affect large scale redistribution of material in the disc, and spiral shocks
can provide suitable locations for accelerated early grain growth
\citep{2004MNRAS.355..543R, 2009MNRAS.398L...6C, 2016MNRAS.458.2676B}
and chemical processing \citep{2011MNRAS.417.2950I} 
. High resolution simulations, with
the capacity to resolve the  large dynamic range of size scales 
associated with `gravito-turbulence' 
will form an essential theoretical counterpart to 
new observational discoveries.  
 
 Turning now to the more abundant lower mass discs which dominate the population
on timescales of Myr, one 
of the most important evolutionary processes (apart from
planet formation itself) is the redistribution/removal of angular momentum from
orbiting dust and gas. Any such process
drives radial flows, redistributing material in the disc and causing accretion on to
the star. Since protoplanetary discs are observed to be accreting at rates which imply
a significant fraction of the disc should be lost to the star over a Myr timescale
\citep{1998ApJ...495..385H,2016A&A...591L...3M}, it is clear that protoplanetary discs should be
considered as {\it accretion discs}. What is not clear, however, is the process
driving this angular momentum transfer.
 A front running mechanism for angular momentum redistribution in recent decades has
been  the
\hbindex{magneto-rotational instability} \citep{1991ApJ...376..214B} a linear instability
of weakly magnetised discs under ideal MHD that operates in any disc in which
the angular velocity decreases outwards.
While this is found to
be effective for moderately ionised conditions \citep{2010ApJ...713...52D,2012MNRAS.422.2685S },
\citet{1996ApJ...462..725G} first pointed out that
finite resistivity should limit the effective operation of the MRI to regions
of suitably high  ionisation level
and that elsewhere (between $\sim 0.3$ and $10-30$au) the
disc has an extensive  MRI `dead zone'; there has subsequently been considerable interest in linking
the low levels of magnetohydrodynamical turbulence in or at the boundaries of such zones  
with conditions conducive to planet formation \citep{2013MNRAS.433.2626R,2016ApJ...816...19H} and to examining how low effective viscosity in such regions
affects the accretion and migration history of planetary cores 
\citep{2007ApJ...660.1609M,2013ApJ...778...78H}.

 Recent years have seen the first attempts to characterise the level of
magnetohydrodynamical turbulence in discs using spatially resolved
observations of molecular line emission   
\citep{2015ApJ...813...99F, 2015A&A...574A.137T}. The analysis involved is highly delicate in  that the
signature of turbulence in line profiles depends on being able to accurately
subtract away the line profile that is expected from thermal
broadening and Keplerian shear alone. Turbulent levels are found to be low
but it is presently unclear whether they contradict the predictions of
MRI generated turbulence.

 Meanwhile, a recent change in direction has been provoked by simulations which include
other non-ideal MHD effects in addition to resistivity. In particular, it has been
found that disc regions that were considered to be beyond the traditional MRI dead zone
are subject to strong damping of magneto-hydrodynamical turbulence
by ambipolar diffusion 
\citep {2015A26A...574A.137T,2013ApJ...775...73S,2013ApJ...772...96B}.
This suppression of the MRI is so effective that
accretion cannot be driven at observable levels unless the disc is threaded by a net vertical field.
In this case, however, it is found that instead of small scale magnetoturbulence driving angular
momentum transport in the disc plane,  angular momentum and mass are  instead removed in the form
of a large scale magnetohydrodynamical wind. Currently, this conclusion is based on local
(shearing box) simulations and a clear goal for the next decade is to establish the reality
or otherwise of such flows in global simulations \citep{1701.04627}.

 {\it If} this picture of large scale magneto-hydrodynamic winds turns  out to be correct then
it will prompt  a paradigm shift concerning our understanding of  secular disc evolution 
and would imply,  for example,
that discs shrink rather than grow with time and that in principle a significant
fraction of disc gas could   be ejected rather than being accreted onto the star.
Currently efforts to test
this scenario observationally are in their infancy and over-lap with efforts to test 
models for  photoevaporation 
(see below).

\begin{figure}
\centering
\includegraphics[scale=0.4]{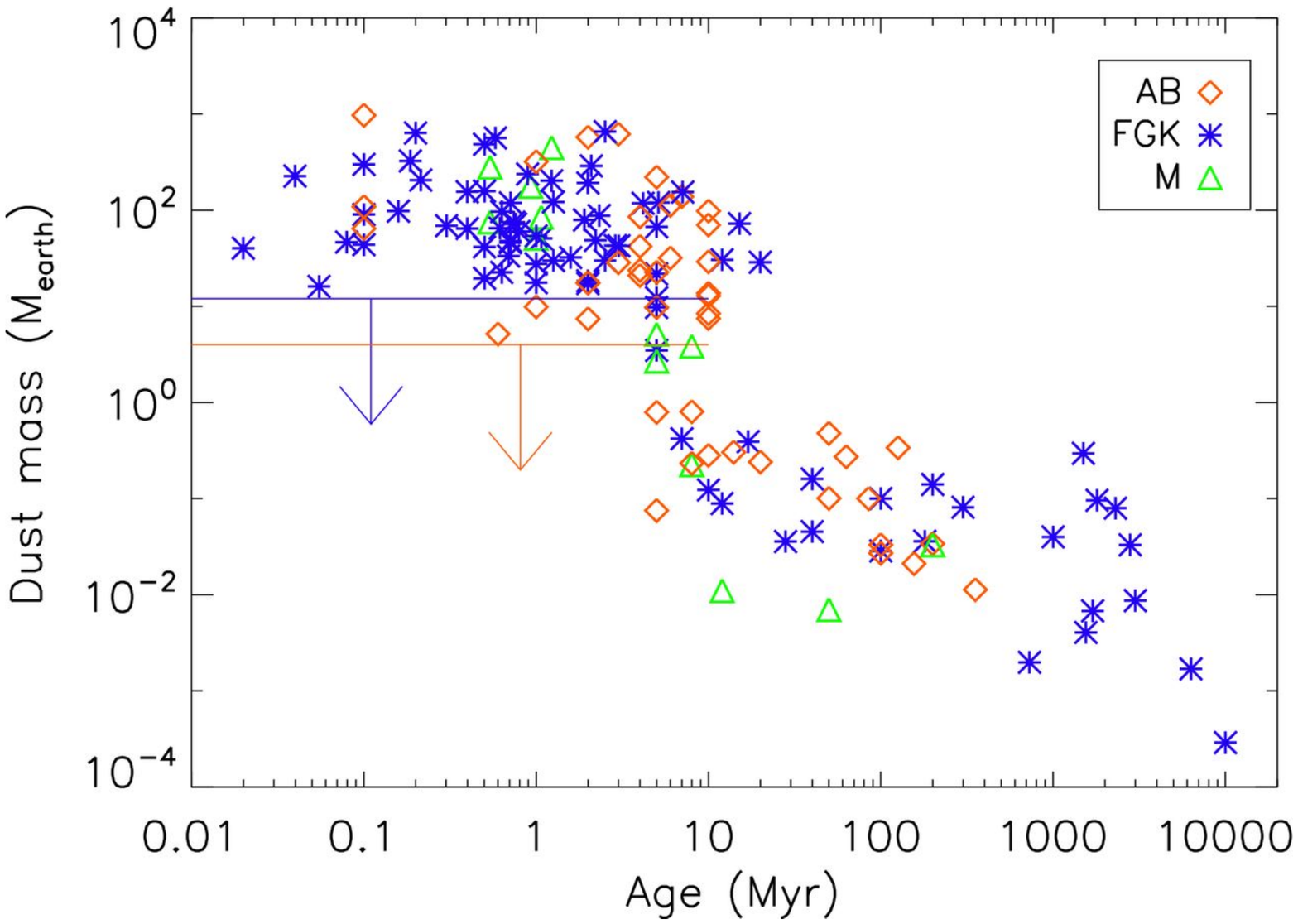}
\caption{The evolution of the mass of dust in mm sized grains from the protoplanetary to the debris disc
phase, with data colour coded according to the spectral type of the central stars. Protoplanetary
and debris discs are well separated in terms of age and dust mass but note the higher upper limits
on detectable dust masses in young systems given the typical distances to star forming clouds \citet{2013MNRAS.435.1037P}.}
\label{fig:gia}
\end{figure}

The disc bearing (Class II) lifetime of protoplanetary discs is typically in the range of
a few Myr \citep{2001ApJ...553L.153H,2010A&A...510A..72F}.  
The subsequent evolutionary stage
(Class III) is compatible, from the point of view of the spectral energy distribution,
with being essentially disc-less - not only is there no evidence for accretion
on to the star, but the lack of near-infrared excess goes hand in hand with undetectably
low levels of far infrared and submm emission \citep{2000A&A...355..165D,2013ApJ...762..100C}.  For solar mass stars, this places upper limits on
the quantity of mm size dust of less than a few earth masses, which is around an
order of magnitude higher than the quantities (see Fig.~\ref{fig:gia}) of  such dust detected in the youngest
debris discs \citep{2013MNRAS.435.1037P}. Likewise, CO is not
detected in non-accreting  T Tauri stars
\citep{2015A&A...583A..66H}, 
implying that the disappearance of disc dust is correlated
with the dispersal of its gas also. 
It is still unclear what processes drive disc
clearing 
(i.e. effect the change from Class II to Class III status) but one thing that has
become obvious is that it cannot be achieved by a simple viscous draining
of material on to the star: an extrapolation of observed disc masses
and accretion rates in Class II sources would imply that they would then
lose their infrared excess over hundreds of Myr and would spend the
majority of that period with the colours of optically thin infrared emission.  
This is contrary to the observational situation \citep{2011MNRAS.410.671E,2013MNRAS.428.3327K}, where discs are either largely
optically thick in the infrared (although with some transition discs evidencing
cleared inner regions in their spectral energy distributions: see earlier) or else
essentially disc-less. Some process, acting on a timescale that is a small fraction
of the typical disc lifetime, is responsible for achieving this final dispersal (see \citealt{2014prpl.conf..475A} for a review of possible
dispersal mechanisms).

\begin{figure}
\centering
\includegraphics[scale=0.3]{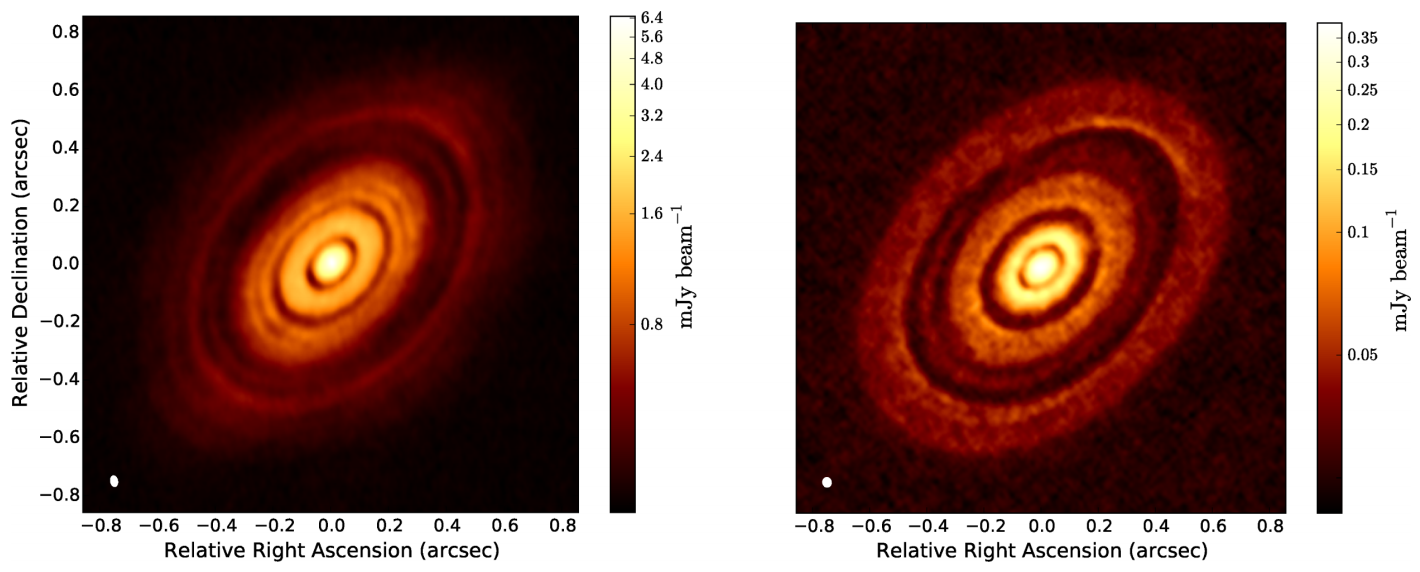}
\caption{This famous ALMA 1.3 mm continuum image (ALMA Partnership 2015) of annular dust rings in \hbindex{HL Tau} (left) is remarkable in that 
its spectrum had given no grounds to suspect the presence of such structure. This raises
the possibility that many young discs may contain such structure. The panel on the right 
is  a simulation by \citet{2015MNRAS.453L..73D}  in which three sub-Jovian mass planets are located at $13,32$ and $69 $au.}
\label{fig:giant}
\end{figure}

 It is currently unclear how transition discs fit into this evolutionary scenario.
At the time that they were first identified, first through anomalous spectral energy
distributions and then subsequently via targeted imaging, they were believed to
represent a minority class, constituting around $10-20 \%$ of all Class II objects.
Objects with such cleared inner regions were thus seen as short-lived immediate 
precursors of disc final clearing. While this picture may still have some merit
it has become complicated by the recent insights provided by  
spectacular images such as the ALMA Science Verification Data on HL Tau
(ALMA Partnership 2015) which shows pronounced annular {\it structures} in a disc (see Fig.~\ref{fig:giant}) which
showed no signature of partial clearing in its spectral energy distribution
and which moreover - from its high accretion rate  -
is thought to be a young system. 
This unexpected evidence of structure 
in a disc not previously identified as a transition disc has opened up the possibility
that the majority of Class II discs may turn out to be similarly structured, in which
case such structure (whether produced by a planet or some other agent)
is not an indication of  imminent disc dispersal. Ongoing  high resolution
imaging  programmes with ALMA based on 
an {\it unbiased} sample of Class II discs 
will do much to clarify the incidence and nature of dusty structures
in protoplanetary discs.  

 Meanwhile there are further ways of exploring the mechanism for disc dispersal through
examining the evidence for {\it  disc winds}. Such winds may represent the
MHD winds described above or else photooevaporative winds driven by ultraviolet/X-ray
radiation, either from the central star or the star forming environment.
The theory of photoevaporation is well developed compared with that of MHD winds 
(see 
\citealt{2001MNRAS.328..485C,2006MNRAS.369..229A,2010MNRAS.401.1415O,2012MNRAS.422.1880O,2015ApJ...804...29G},
in the
case of evaporation by the host star and 
\citealt{1998ApJ...499..758J,2004ApJ...611..360A,2016MNRAS.457.3593F}
for photoevaporation driven by
neighbouring higher mass stars).
Photoevaporative winds 
are predicted to be a significant sink of mass at radii
beyond a few au and to be important agents of dispersing the last remnants of disc gas 
at late evolutionary stages of protostellar disc evolution.
 The narrow components of a number
of optical and near-infrared lines in protostellar discs (such as NeII and OI) can
be explained by photoevaporation models \citep {2008MNRAS.391L..64A,2010MNRAS.406.1533E}  . 
To date there has been no similar exploration of the observability of MHD winds,
in part on account of the lack of self-consistent global models and in part because
of the observational difficulty of disentangling the signatures of such winds
(which can launch from the inner disc at hundreds of km s$^{-1}$ ) from that of
similarly high velocity outflows associated with jets.

 Mapping at cm wavelengths using the VLA or e-Merlin  can potentially
provide observational constraints on the rate of  mass loss in ionised flows
since imaging in the free-free continuum can be used to map the
distribution of extended  ionised  gas around protostellar discs. Currently the resolution
attainable (10s of au) offers the possibility of distinguishing between
photoevaporative winds driven by EUV radiation and the denser conditions
produced in the more vigorous X-ray driven winds. Such 
observations will also
be confronted with the predictions of global MHD wind models as these
become available.

\section{Debris discs}

\begin{svgraybox}
\hbindex{Debris discs} are the circumstellar discs found around main sequence stars. They are made up of asteroids, comets, dust and gas, all of which may be interspersed within a planetary system.
Planetesimals in these systems collide and are ground down to dust that is readily detectable through the infrared (IR) excess it creates.
Thousands of such debris discs are known and for around a hundred, we have been able to make a resolved image of these discs.
Many of these planetesimal belts are cold ($T \lesssim 100$K) and observed in the far-infrared (as such, they may be considered analogues to the Kuiper belt in our Solar System).
However, dust very close to its host star ($T \gtrsim 300$K, analogous to the Solar System's zodiacal cloud) is also observed around a significant fraction of stars.
In addition, gas is detected in a growing number of debris discs.
Moreover, dust and gas are also observed around the oldest stars that had time to transform into white dwarfs.
\end{svgraybox}


Far-IR emission from debris disc dust is found around stars of all spectral type.
Detection rates are around 20\% for A-K spectral types, with some evidence for a fall-off in rate toward later spectral types \citep{2013A&A...555A..11E,2014MNRAS.445.2558T}.
The debris discs that are detected are much more massive than our Solar System's Kuiper belt. 
There are few dust detections for M-stars, but this does not mean that M stars do not have discs, since the low luminosity of such stars reduces the detectability of any emission in the current surveys.
The improved sensitivity of future far-IR missions (like \hbindex{SPICA}) offers the potential to discover more discs around these late-type stars.
The paradigm to explain the observed cold dust emission is that the dust is produced from a reservoir of big planetesimals that slowly depletes and grinds down to dust in a process known as a collisional cascade.
This is supported by the (on average) lower infrared emission from older stars.
The size of the biggest bodies composing the belt is not well-known but should be large enough (at least a few km in diameter) for the belt to collisionally survive for billions of years as observed \citep[i.e. $\gtrsim 10$km,][]{2008ApJ...673.1123L}.
These discs can be considered as the left-overs of the planet formation process, most of which occurred in the protoplanetary disc phase described in the previous section.
As such debris discs provide information on the outcome of the planetesimal and planet formation processes that went on at earlier epochs.

\subsection{Birth of debris discs}\label{birth}
There are still many unknowns as to the origin of debris discs and the different steps from the protoplanetary disc phase (see previous section) leading to their creation \citep[see][for more detail]{2015Ap&SS.357..103W}.
For example, it is not yet clear if the planetesimals that replenish the dust seen in debris discs are already present early on in the protoplanetary disc phase (e.g. in the rings observed in HL Tau or TW Hydrae, see Fig.~\ref{fig:giant}).
Our ignorance is driven by the difficulty of detecting the planetesimals, but also by a lack of understanding of how to overcome the bouncing barrier \citep{1993Icar..106..151B} and radial drift \citep{1977MNRAS.180...57W} that otherwise prevent sub-$\mu$m sized dust in
protoplanetary discs growing beyond cm in size. 
One solution is that planetesimals form through gravitational instabilities in dense regions of the protoplanetary disc, perhaps at locations where dust density has been enhanced by the \hbindex{streaming instability} \citep[an instability in which dust grains concentrate into clumps owing to gas drag leading to
their gravitational collapse,][]{2010AREPS..38..493C,2014prpl.conf..547J}.
This may mean that planetesimals form at favoured locations in the protoplanetary disc \citep[e.g., near snow-lines,][]{2017arXiv170202151S}, or just outside the gaps carved by planets or in spirals of gravitationally unstable discs (see Fig.~\ref{figgi}).
However, since the streaming instability is enhanced when the gas is depleted relative to the dust \citep{2015A&A...579A..43C}, it is possible that planetesimals are preferentially formed late on, while the disc is in the process of having its gas dispersed by photoevaporation \citep{2017Carrera}.

\begin{figure}
\centering
\includegraphics[scale=0.3]{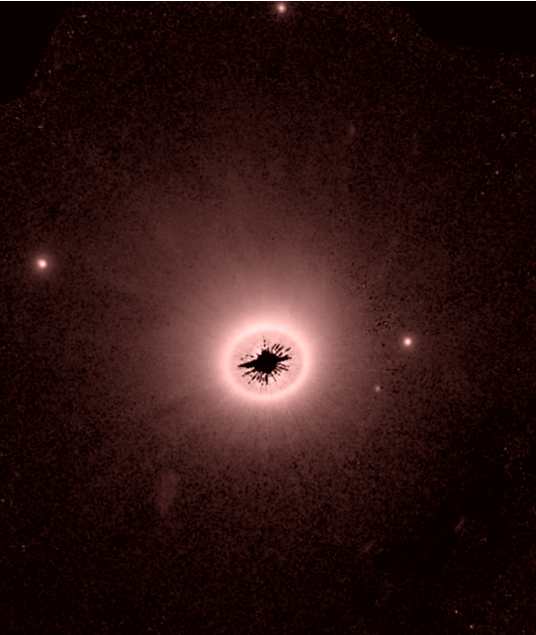}
\includegraphics[scale=0.26]{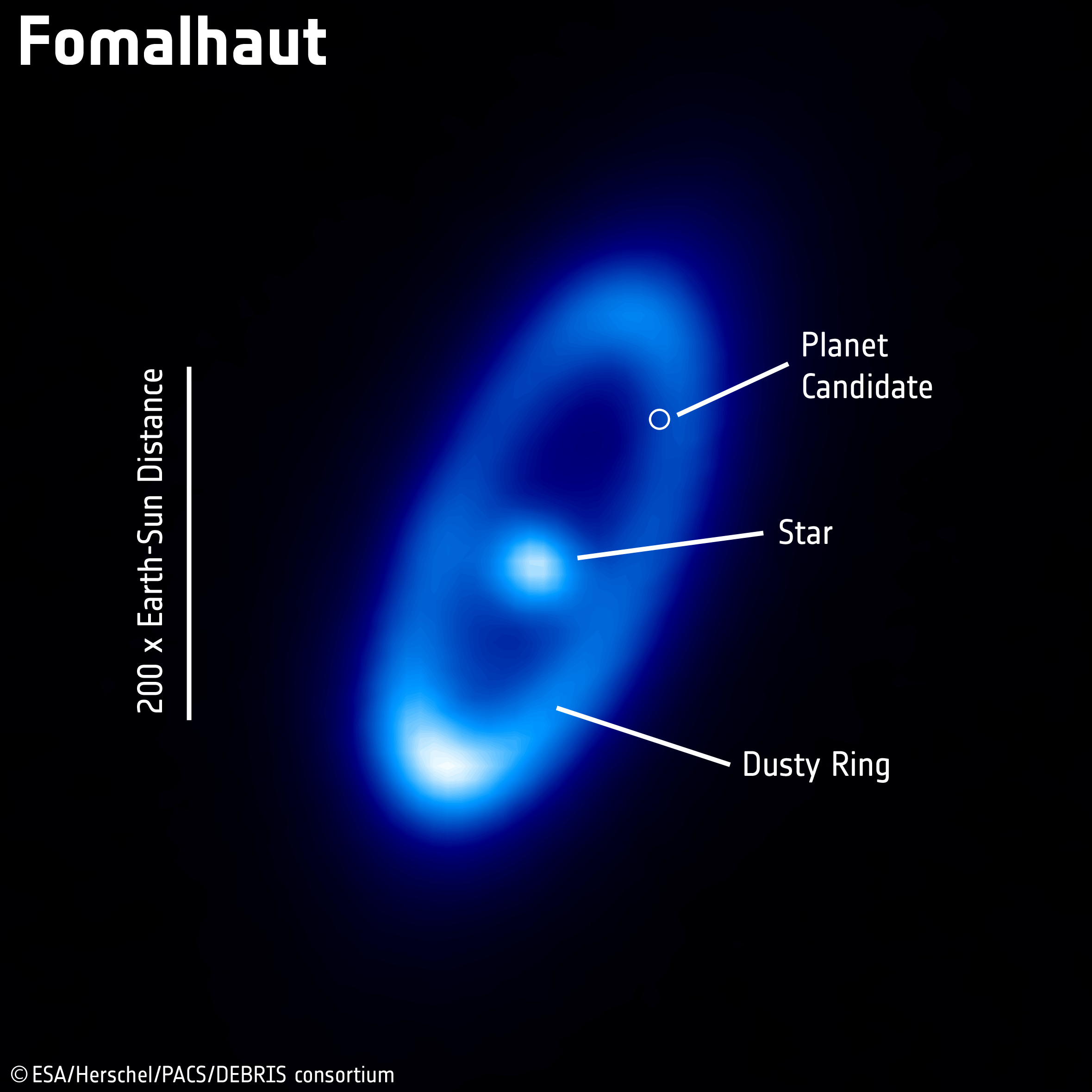}
\caption{Two ring-like debris discs: {\it left)} HD 181327 observed with HST \citep{2014AJ....148...59S}; {\it right)} Fomalhaut observed with Herschel \citep{2012A&A...540A.125A}.}
\label{fig:1}
\end{figure}

These different possibilities make different predictions for the radial location and radial width of the region in which planetesimals would be expected to form which can be compared with the observed properties of debris discs.
Studies of nearby debris discs around main sequence stars with a range of ages tell us that planetesimal belts can be present at up to $\sim 200$au, but the known debris belts are more commonly closer in at $\sim 40$au.
They are also often in belts that are radially narrow $dr/r=0.1$ (e.g., \hbindex{HR 4796}, \hbindex{HD 181327}, \hbindex{Fomalhaut}; see Fig.~\ref{fig:1}), although there are examples of broad belts too $dr/r>1$ (e.g. $\beta$ Pic, see Fig.~\ref{fig:betapic}), and systems with belts at multiple radii \citep[e.g.][]{2015ApJ...798..124R}.
These observations already provide distributions that can be used to constrain models of planetesimal formation, though such comparisons are only just beginning to be made.
However, high resolution imaging of a larger sample of debris discs with ALMA, and later with the coming radio array SKA,
will provide better constraints on the location and width of the planetesimal belts in these systems (which is usually inferred indirectly from the emission spectrum or shorter wavelength data).

Studies of young associations (e.g., the TW Hydra association, or the $\beta$ Pic moving group) allow to probe the properties of debris discs straight after their formation.
These show a diversity of disc radii and widths that is not significantly different to that of
the discs of older main sequence stars.
Thus, there is no evidence for an increase in disc radius with age \citep{2005ApJ...635..625N} as expected in models in which a debris
belt only becomes sufficiently luminous to be detectable once sufficient time has elapsed for planetesimals
to grow into Pluto-sized objects \citep[which takes longer further from the star,][]{2008ApJS..179..451K}.
However, there are relatively few debris discs known at this early epoch, and further studies of debris discs close to the transition are needed to assess if there is any evolution other than the decrease in brightness expected from collisional grinding.
With the help of ALMA and future potential missions such as \hbindex{SPICA}, or a 10m-aperture \hbindex{far-IR surveyor} (such as the one proposed by NASA called the Origins Survey Telescope) it will be possible to identify and characterise debris disc dust levels in nearby star forming regions.

The evolution of the debris in the transition phase is poorly constrained at present, yet the
dynamics of this transition can result in observable (i.e., testable) phenomena.
For example, the dispersal of the protoplanetary disc may sweep the remaining mm-cm sized dust into belts \citep{2007MNRAS.375..500A} that, assuming this mass does not coalesce into planetesimals, would be both luminous and short-lived.
Sweeping can also occur through interaction with planets that formed closer in, since these may undergo
migration shortly after formation \citep{2003ApJ...598.1321W,2011Icar..211..819C} or be scattered into an outer planetesimal belt.
Bright rings of $\mu$m-sized dust can also be created without in situ planetesimals through the action of gas drag on such small dust \citep{2001ApJ...557..990T}.
This simply requires an inner planetesimal belt and a substantial gas disc, and may possibly explain the two narrow rings seen in scattered light at 100s of au in HD 141569 despite mm-sized grains not being detected
at these locations \citep{2016ApJ...829....6W}. 
Again, further observations of systems in the late phases of protoplanetary disc evolution, or in the early stages of debris disc formation, will help to understand this transition.

Another property of the known debris discs is that their planetesimals must collide at high enough velocities to create significant quantities of dust.
Since collision velocities are expected to be low in a protoplanetary disc due to damping by gas drag,
there must be some process that stirs the planetesimal belt.
Possibilities include that the planetesimals inherit a large velocity dispersion from their formation process \citep[e.g.][]{2013MNRAS.431.1903W}, or from some other (as yet undefined) aspect of protoplanetary disc evolution, or dispersal causes them to end up with a large velocity dispersion 
as soon as the protoplanetary disc has dispersed. \hbindex{Stirring} could also occur after protoplanetary disc dispersal; e.g. the planetesimals could be born with a low velocity dispersion resulting in their growth into Pluto-sized objects that stir the disc \citep[i.e., self-stirring,][]{2001AJ....121..538K,2010MNRAS.405.1253K}, or an interior planetary system could stir the disc \citep{2009MNRAS.399.1403M}.
Population studies of debris discs are inconclusive as to the origin of the stirring, but detailed investigations of individual systems allow constraints to be set within the context of the different scenarios on, say, the mass and orbit of the perturbing planet, or the surface density and initial planetesimal sizes for a self-stirred disc.

Such detailed studies of individual discs can also provide information on the level of stirring.
For some systems the vertical height of edge-on discs suggests a low level of stirring \citep[$<5$\%, e.g. AU Mic,][]{2005AJ....129.1008K}, while a sharp outer edge in others suggests likewise \citep{2008A&A...481..713T}. 
Herschel also discovered a population of \hbindex{cold debris discs} \citep{2013A&A...555A..11E} that are best explained as unstirred debris belts of planetesimals that are 10s of metres in size that evolve very slowly but producing low quantities of dust \citep{2013ApJ...772...32K}.
However, the possibility that the emission from these cold discs arises from background galaxies still needs to be unambiguously ruled out \citep{2014ApJ...784...33G}, perhaps by confirming that the emission is co-moving with the stars in question.
Overall, high resolution studies of larger numbers of discs are required to determine the stirring level and its dependence on other properties of the system.


\subsection{Links between debris and planets}
\begin{figure}
\centering
\includegraphics[scale=0.22]{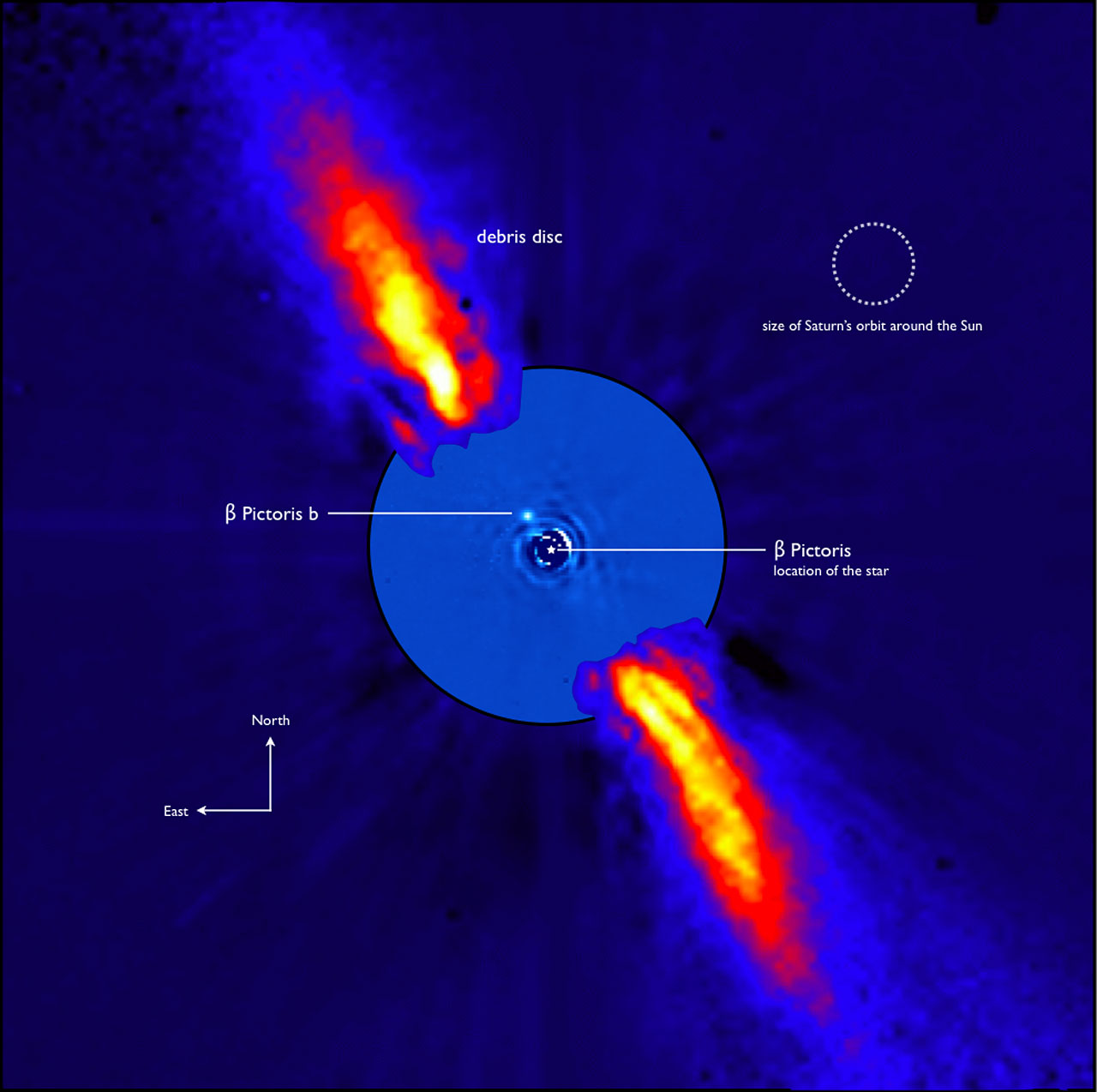}
\caption{$\beta$ Pic's debris disc and planet. This is a composite image in the near-infrared for which the ADONIS instrument on ESO's 3.6m telescope was used to observe the outer part of the disc and the inner part is seen using the NACO instrument on the VLT. 
The detected planet $\beta$ Pic b is at a projected distance of $\sim$ 9au and has a mass about 7 times that of Jupiter. ESO/A.-M. Lagrange et al.}
\label{fig:betapic}
\end{figure}

There are currently $\sim 40$ systems known to host both a debris disc and a \hbindex{planet} \citep{2014A&A...565A..15M,2015ApJ...801..143M}.
Since both debris discs and planets are thought to form in protoplanetary discs, it might be expected that the properties of these two components should be correlated somehow, for example because the protoplanetary disc properties that are favourable for forming planets might also be conducive for the formation of planetesimals.
However, early studies found no such correlation \citep[e.g.][]{2009ApJ...705.1226B}, which was attributed to the fact that the known debris discs are typically located at a few 10's of au, while planets are typically much closer in.
More recently tentative evidence has been found for a correlation between the presence of low-mass planets detected in radial velocity surveys and debris \citep{2012MNRAS.424.1206W,2014A&A...565A..15M}, and a possible anti-correlation with giant planets which could more easily scatter any debris that would remain \citep{2012A&A...541A..11R}.
It is also notable that most of the systems with planets that have been directly imaged at $\gg$ 5au also have debris discs, the most famous being \hbindex{$\beta$ Pic} (see Fig.~\ref{fig:betapic}) with a $\sim 7$M$_{\rm Jup}$ planet at $\sim$ 9au \citep{2013ApJ...776...15C}, 
\hbindex{HR 8799} with 4 detected planets sandwiched between inner and outer debris belts \citep{2008Sci...322.1348M,2010Natur.468.1080M,2014ApJ...780...97M,2016MNRAS.460L..10B}, and \hbindex{Fomalhaut} with a narrow debris ring and an eccentric planet \citep{2013ApJ...775...56K}.

Our understanding of the links between debris and planets could be extended on two fronts, by using statistical methods to identify correlations between the properties of a system's debris disc and its planetary system, and by using resolved images of debris discs to characterise the way in which planets interact with debris and to identify planets that could not have been detected otherwise (and to learn about their formation history).
The current statistical studies mentioned above are limited by the small number of detections, both of debris and of planets.
This will improve significantly with future far-IR missions that can detect debris around more distant stars (and so increase the number of known debris discs) as well as detect debris discs down to fainter levels approaching that of the Solar System's Kuiper belt.
A variety of exoplanet detection techniques will also increase the number of known exoplanets.
For example, \hbindex{Gaia} is expected to discover at least 20,000 planets (more massive than Jupiter with periods out to several au) within 500pc \citep{2014ApJ...797...14P}, meaning that 1000's should be discovered in systems with a bright debris disc, the closest of which will be known from Herschel surveys (out to 20-50\,pc), a distance limit which can be extended in future far-IR surveys.
Upcoming transit missions will also be able to detect small Earth-like planets in edge-on systems: \hbindex{TESS} is expected to detect $\sim$ 2000 planets \citep[among which 500 will have a radius smaller than twice the Earth,][]{2015ApJ...809...77S}, and \hbindex{PLATO} is expected to detect 100's 
of earth-like planets and 1000's greater than Neptune \citep{2014ExA....38..249R}.
Some of these systems will have constraints on their debris disc from far-IR surveys, but these
are also good candidates to search for correlations with hot dust co-located with the planets (see next section).

If there are planets in a system hosting a debris disc then it is inevitable that the planets' gravitational perturbations will impose structure on the debris disc. 
It is known that planet-disc interactions can create detectable asymmetries: clumps \citep{2003ApJ...598.1321W}, offsets \citep{1999ApJ...527..918W}, warps \citep{2001A&A...370..447A}, gaps \citep{2016MNRAS.462L.116S}, spirals \citep{2005A&A...440..937W}.
Our understanding of the variety of features that planets can produce is also growing; e.g., planets can also create \hbindex{scattered-disc analogues} (similar to the one in our Solar System mostly populated by icy minor planets at $>$ 30au) and mini-Oort clouds that are potential outcomes of planetary systems \citep[see][]{2017MNRAS.464.3385W}.  
Thus observations of such features in debris disc images can provide evidence for planets that would otherwise be undetectable.
For instance, in $\beta$ Pic (see Fig.~\ref{fig:betapic}), the detected planet was first hypothesised because of the observation of a \hbindex{warp} in the $\beta$ Pic dust disc \citep{1997MNRAS.292..896M}.
Another example is the growing number of detections of eccentric discs, which is thought to be due to the presence of eccentric planets secularly forcing the disc to become eccentric over long timescales \citep{1999ApJ...527..918W,2016ApJ...827..125L}.

In the coming years the number of features observed in debris discs will grow as new instrumentation becomes available.
ALMA is already providing images of structures in the parent planetesimal belts of some systems.
The small inner working angles, high-resolution and contrast of new instruments such as \hbindex{SPHERE} or \hbindex{GPI} are also providing scattered light images that reveal new structures not foreseen by models with no easy interpretation \citep[e.g. AU Mic,][]{2015Natur.526..230B}.
Scattered light imaging capability will continue to improve with \hbindex{JWST}, \hbindex{WFIRST} and the \hbindex{ELT}, and thermal imaging capabilities will improve with ALMA, JWST and SKA.
A multi-wavelength approach is particularly crucial to test models for the origin of a given structure, since the interpretation of a given feature is often degenerate \citep[e.g.][]{2006ApJ...639.1153W}.
For example, a dust and gas clump like that seen around $\beta$ Pic \citep{2014Sci...343.1490D,2017MNRAS.464.1415M} can arise from resonances with a planet or from a single massive collision at 10's of au \citep{2014MNRAS.440.3757J, 2015A&A...573A..39K}, although in this case the breadth of the gas clump rules out a giant impact origin \citep{2017MNRAS.464.1415M}.  

Unambiguously identifying debris disc structures with known planets is important to test and further refine our understanding
of planet-disc interactions \citep[e.g.][]{2012A&A...547A..92T} and the processes of their formation and evolution.
This gives a much better handle on the origin of these structures (linked to planets or not) and allows us to better constrain debris disc models and refine some of the physics used in these models.
While this is challenging because the planets are often hard to find with other methods, the brightest planets can be detected in direct imaging, the capabilities for which are improving with the same instrumentation used to image the disc (i.e., JWST, WFIRST, ELT). 
Even if detections of low mass planets (i.e., Neptunes) in the outer regions (i.e., $>5$\,au) of specific systems may remain challenging, our understanding of the frequency of such planets will be transformed by the microlensing surveys of EUCLID and WFIRST, and this will significantly inform our interpretation of debris disc structures.

\subsection{Debris in the middle of planetary systems}
While most debris discs are made up of a cold belt at 10's of au, we know of the existence of many two-temperature debris discs that are mainly probing systems with multiple belts such as the Kuiper belt and the Asteroid belt in our Solar System \citep{2014MNRAS.444.3164K}.
Dust within a few au of its host star is also observed around a large fraction of systems irrespective of the existence of a cold outer belt.
When this dust is warmer than around 300K it is referred to as an ``\hbindex{exozodi}'' in reference to the \hbindex{zodiacal dust} in our Solar System that surrounds the innermost planets and goes all the way to a few solar radii.
We distinguish hot dust (up to $\sim$ 2000K, very close to the host star) and warm dust ($\sim$ 300K, in the habitable zone of the system) from an observational perspective as the former is observed in the near-IR and the latter in the mid-IR.
Current near-infrared interferometry studies have detected hot dust around $>10$\% of stars \citep{2014A&A...570A.128E}, with surprisingly little dependence on the properties of the host star or its outer debris belt.
Mid-infrared photometry has shown that bright warm dust (brighter than around 10\% of the stellar photospheric level at 12$\mu$m) is relatively rare around old nearby stars but more common around young stars \citep{2014MNRAS.444.3164K}.
However, mid-infrared interferometric techniques show that lower levels of dust (at the 0.1\% above photospheric level with the Keck Interferometer Nuller) may correlate with the presence of an outer debris belt \citep{2014ApJ...797..119M}.

\begin{figure}
\centering
\includegraphics[scale=0.35]{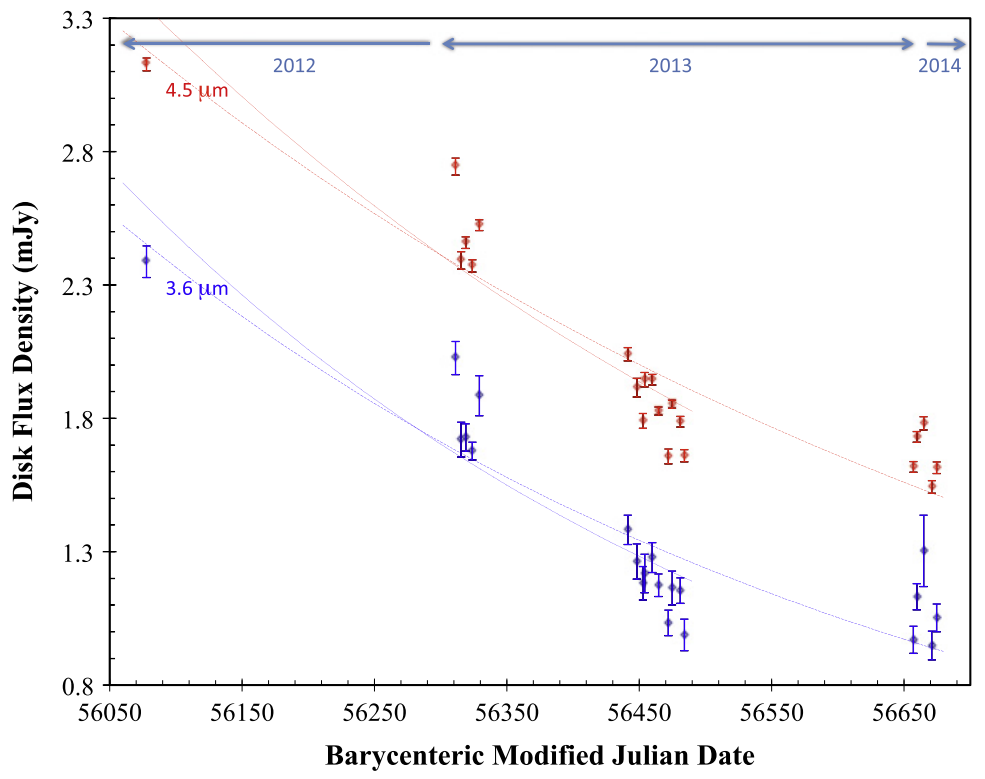}
\caption{Flux decrease at 4.5 (red) and 3.6$\mu$m (blue) in the P1121 debris disc over more than 2 years. The solid lines are fit to the data assuming an exponential decay \citep{2015ApJ...805...77M}.}
\label{fig:2}
\end{figure}

The origin of exozodi dust is uncertain at present \citep[see the review by][for more details]{Kra17}.
The high luminosity and temperature of the hot dust defies easy explanation, because its collisional depletion at its inferred proximity to the star prevents its accumulation.
One of the proposed explanations involves magnetic fields trapping nano-grains \citep{2016ApJ...816...50R}, underlining that the physics in these highly collisional and hot systems may vary from typical colder belts.
Warm dust that can be at larger distance from the star is easier to explain.
For young stars ($<100$\,Myr) with dust within a few au, the favoured explanation is in a \hbindex{massive collision}, which like the Moon forming collision with Earth is expected in the late stages of planetary formation \citep[e.g.][]{2006AJ....131.1837K,2009Icar..203..644R,2012MNRAS.425..657J}.  
Thus warm dust detections provide a way of probing on-going planetary formation \citep[e.g.][]{2009ApJ...701.2019L}, an interpretation which is supported by photometric variations (on $<$1yr timescale) in the infrared \citep[see Fig.~\ref{fig:2},][]{2014Sci...345.1032M,2015ApJ...805...77M}.
However, for older stars ($>100$\,Myr), such giant impacts are expected to be rare, and collisional depletion precludes detectable levels of warm dust having its origin in an in situ asteroid belt \citep{2007ApJ...658..569W}.
Instead the observed warm dust could be supplied from an outer planetesimal belt, either through Poynting-Robertson drag that transports the dust inward \citep[][]{2014A&A...571A..51V,2015MNRAS.449.2304K}, or through the scattering in of planetesimals by interaction with planets resulting in exocomet activity \citep{2014MNRAS.441.2380B,2016arXiv161102196F}.
These possibilities are supported by the aforementioned correlation of bright mid-infrared excesses with outer belts.
For warm dust that is sufficiently far from the star, these can be explained as extrasolar analogues to the asteroid belt \citep[e.g.][]{2017arXiv170205966G}.
A new generation of models \citep[e.g.,][]{2013A&A...558A.121K}, along with the new observations of this phenomenon (discussed below), will eventually unravel the origin of these excesses.

\begin{figure}
\centering
\includegraphics[scale=0.42]{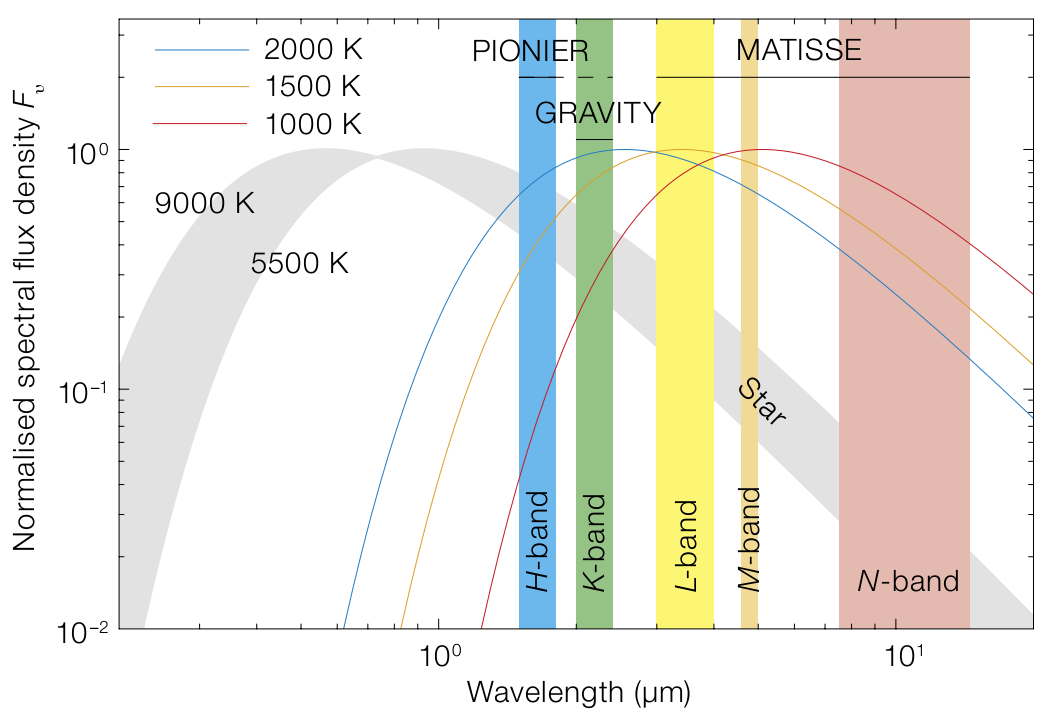}
\caption{Wavelength coverage of the second-generation VLTI instruments (and PIONIER) compared to the
wavelength range in which blackbody dust emission from hot exozodiacal dust peaks \citep{2015Msngr.159...24E}.}
\label{fig:newinstru}
\end{figure}

Regardless of its origin, this dust is a valuable probe of the innermost regions of planetary systems where habitable planets are located.
However, it is also a potential hindrance to the direct detection of Earth-like planets in the habitable zone; e.g., even a small amount of dust (10-20 times brighter than our faint zodiacal cloud) could severely hamper detections \citep{2010A&A...509A...9D}.
As such, there is currently much effort in trying to characterise dust in the habitable zone of nearby stars down to much lower levels; e.g., the NASA HOSTS survey is using the LBTI nulling interferometer to search the nearest stars for mid-infrared excesses down to 0.05\% above the photospheric level \citep{2016ApJ...824...66D}, with the aim of identifying exozodi-free targets that are most suitable for future searches for Earth-like planets.
While LBTI provides limited information on the spatial distribution of any exozodi that it finds, it may provide the first hints of the clumpy structures expected as dust interacts with an inner planetary system \citep{2015MNRAS.448..684S}.
JWST also has the potential to detect the asymmetric dust distribution expected to persist for Myr from giant impact debris \citep{2015A&A...573A..39K}, and to search for any photometric variability.
Future direct imaging missions such as WFIRST will also provide the very small inner working angle (0.1''), high resolution and constrast needed to resolve these \hbindex{Solar System analogues}, a capability which will be further improved upon by ELT.
Second generation VLTI instruments such as \hbindex{GRAVITY} and \hbindex{MATISSE} (together with the current \hbindex{PIONIER}) will lead to multi-wavelength measurements over a large range of dust temperatures (see Fig.~\ref{fig:newinstru}) 
that may reveal a connection between the hot and warm dust and will explore dust properties through the potential detection of spectral features \citep[e.g. the 3 and 10$\mu$m silicate features,][]{2012A&A...541A.148E}.
VLTI in the southern hemisphere complements the \hbindex{LBTI} in the north but is not as sensitive.
A new concept instrument on the \hbindex{VLTI} using nulling interferometry (to get rid of the stellar contribution) may be built in the future to improve the high constrast capabilities needed to detect these excesses.
In the distant future, improved constraints would be possible with space-based infrared interferometry \citep[e.g.][]{2004AdSpR..34..613F,2016EAS....78...45L}.

Other novel ways are also being proposed to probe hot and warm dust.
For example, small dust clumps embedded in exozodis whose IR-excesses cannot be detected by current instruments can mimic an Earth-like \hbindex{astrometric} signal \citep{2016A&A...592A..39K}, though can be distinguished from a planetary signal by multi-wavelength observations.
Variability on timescales of at least as short as one year has also been found for some hot exozodis \citep{2016A&A...595A..44E}; characterising this variability is the first step to understanding its origin. Polarisation measurements may also be used to find new exozodis \citep[e.g.][]{2016ApJ...825..124M}.
The detection of falling evaporating bodies (through high-velocity gas absorption lines) may also be a good tracer for the presence of hot dust \citep{2000Icar..143..170B,2015AdAst2015E..26W,2016A&A...594L...1E}.
Dips and dimming in the light-curves of nearby stars, such as that already detected by Kepler \citep[][]{2016MNRAS.457.3988B}, which may be detected more commonly by PLATO, may have its origin in transits of exocomets in front of the star, thus providing 
another way of probing planetesimal and dust in the inner regions of planetary systems.

\subsection{Gas in debris discs}

\hbindex{Gas} has been discovered around a growing number of main sequence stars, old enough for their protoplanetary gas discs to have dispersed via accretion, photoevaporation or MHD winds (see previous section on evolutionary processes in protoplanetary discs).
Molecular \hbindex{CO} has been detected in the submm in more than 10 debris discs and is often co-located with its planetesimal belt \citep[e.g.][]{1995Natur.373..494Z,2012ApJ...758...77Z,2015ApJ...814...42M,2016MNRAS.461.3910G,2017MNRAS.464.1415M}.
Atomic carbon and oxygen \citep[e.g.][]{2014ApJ...796L..11R}, as well as metals \citep{2012A&A...544A.134N,2017MNRAS.466.3582H}, have also been detected in a few systems and seem more extended than CO.
Atoms have been detected both with the space-based telescopes HST and FUSE through UV absorption lines \citep{2000ApJ...538..904R,2006Natur.441..724R}, with Herschel through far-infrared emission lines \citep{2014A&A...563A..66C,2016A&A...591A..27B}, and
in the submm \citep{2017arXiv170306661H}.
Most of these observations can be explained by a model in which CO is produced from volatile-rich solid bodies located in the debris belts \citep[as first proposed by][]{2011ApJ...740L...7M,2012ApJ...758...77Z}, which then \hbindex{photodissociates} quickly into C and O that evolve by viscous spreading \citep{2013ApJ...762..114X,2016MNRAS.461..845K,2017Kral}. 
The implication is that for most (but not all) systems there is no requirement for primordial gas to be retained from the protoplanetary disc phase, and any debris disc with icy planetesimals will create second generation CO, C and O gas at some level.
As such, measurement of these gaseous components provides a way to infer the composition of the exocomets from which the gas was created \citep{2015MNRAS.447.3936M,2016MNRAS.461..845K,2016MNRAS.460.2933M,Matra17}.

\begin{figure}
\centering
\includegraphics[scale=0.2]{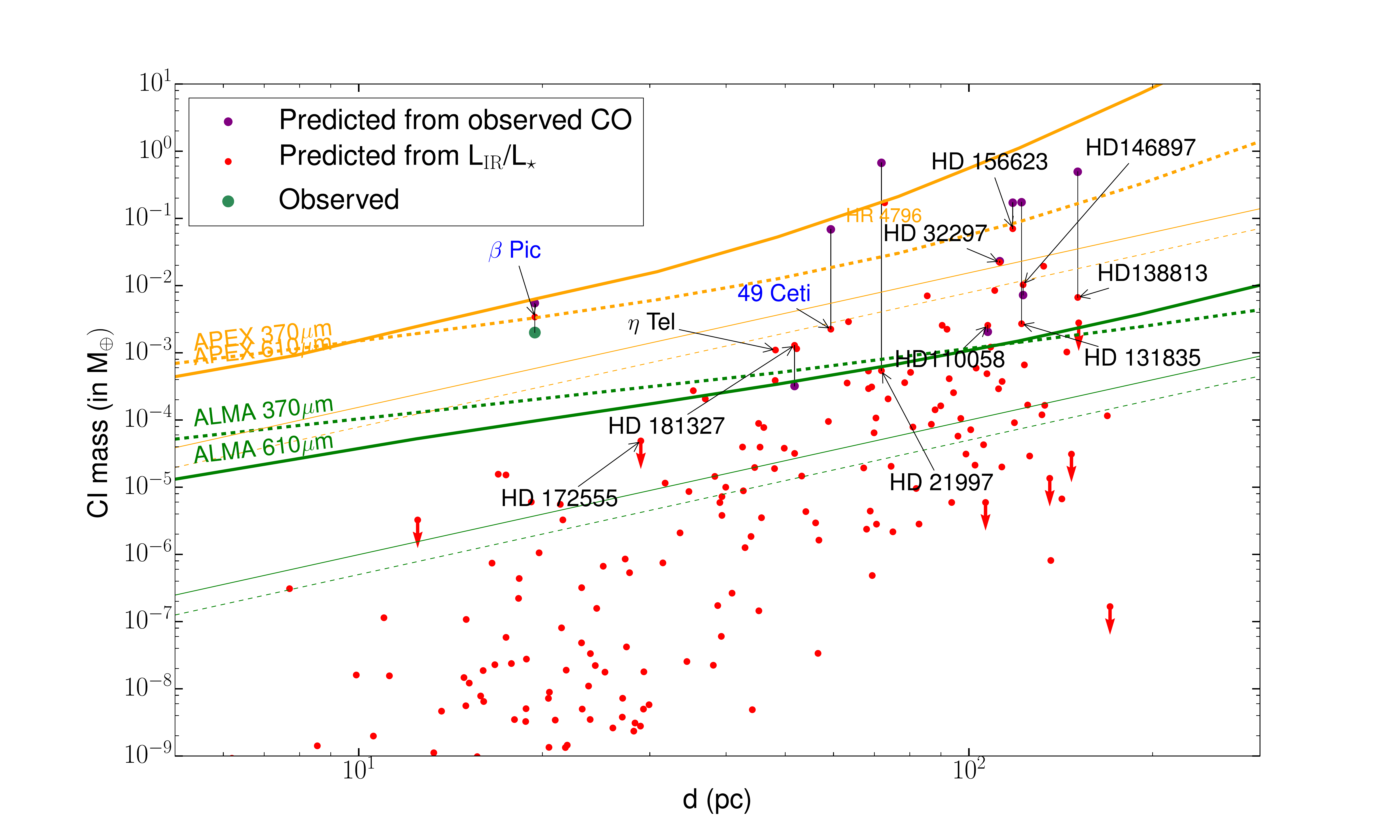}
\caption{CI mass (in M$_\oplus$) as a function of distance to Earth ($d$) from the secondary gas model of \citet{2017Kral}.
Planetary systems with gas detections are labelled with their names.
If CI is already detected the label is in blue (black otherwise).
The CI mass for $\beta$ Pic derived from \citet{2016MNRAS.461..845K} is shown as a green point.
The red points are predictions from the model.
The red downward arrows show systems with grains that are warmer than 140K, which may not be able to keep CO trapped on solid bodies.
The purple points show predictions from the model when the observed CO mass is used rather than the CO mass predicted from $L_{\rm IR}/L_\star$.
Detection limits at 5$\sigma$ in one hour are shown for APEX (in orange) and ALMA (in green) at 370 (dashed) and 610$\mu$m (solid).
The thin lines are for LTE calculations and thick lines for more realistic NLTE calculations \citep[see details in][]{2017Kral}.}
\label{fig:CI}
\end{figure}

To understand the exact dynamics of this gas, its origin, and what we learn from its observations for the planetary system as a whole, a bigger sample of gas detections is required.
In the immediate future, \hbindex{ALMA} is predicted to provide at least 15 new CO detections and 30 new CI detections (see Fig.~\ref{fig:CI}).
CI observations are particularly promising, since in addition to being more readily detected than other species, this component spreads viscously all the way to the star potentially enabling resolved imaging of gas disc structures created by close in giant planets.
The next generation of far-IR missions, such as SPICA, are predicted to enable detection of at least 25 new systems in CII and $\sim$15 in OI \citep{2017Kral},
while a 10m far-IR space telescope could result in $>100$ CII gas disc detections, $>50$ systems with OI detected, some of which potentially resolved.
The expected level of OI gas depends strongly on the amount of water released together with CO from the planetesimals, since this provides extra oxygen in the gas disc from the photodissociation of H$_2$O.
This illustrates how gas observations can provide an estimate of the CO/H$_2$O ratio on exocomets that could be compared to Solar System values (i.e., leading to the taxonomy of exocomets).
The secondary gas generation process is also expected to create hydrogen \citep[][]{2016MNRAS.461..845K}, accretion of which onto the star has been confirmed observationally for $\beta$ Pic \citep{2016arXiv161200848W}, with similar detections possible in other systems.
The combination of CO and \hbindex{CI} or \hbindex{CII} detections will lead to a better understanding of the gas dynamics in these discs, by providing estimates of the disc viscosity and ionisation fraction with which to test theories for how angular momentum is transported in the discs \citep[e.g.][]{2016MNRAS.461.1614K}.

$\beta$ Pictoris is so far the only system for which we have spatially resolved images (using VLT/\hbindex{UVES}) of the metals \citep[such as NaI, FeI and CaII,][]{2004A&A...413..681B}, which are shown to extend inwards to at least $\sim$10au. The high angular and spectral resolution of UVES should be
used on other targets in the future to detect more systems with gas. This technique was first used in \citet{2001ApJ...563L..77O} and the model by \citet{2010ApJ...720..923Z} can be used to make predictions of the different emission line fluxes expected. 
These new gaseous systems could then be followed-up with more detailed UVES/\hbindex{CRIRES} or ALMA observations. This will enhance our understanding of the origin of metals and how they
dynamically evolve in these gas discs.

Other novel ways of observing gas around main sequence stars could be through detecting rovibrational CO lines with the JWST as was already done from the ground for $\beta$ Pic \citep{2011ApJ...738...12T}. HI or OH may also be detectable with future radio telescopes such 
as \hbindex{SKA} \citep{2013arXiv1301.4124A} or the next generation VLA \citep{2015arXiv151006438C}. For systems that are edge-on, using UV absorption lines \citep[e.g.][]{2000ApJ...538..904R} could also enable us to detect new systems with gas. Systems showing the presence of falling evaporating bodies (similar to comets on sun-grazing orbits) through high-velocity gas absorption lines
\citep[e.g.][]{2014Natur.514..462K,2014A&A...561L..10K,2012PASP..124.1042M} should be edge-on and may be promising targets for future UV absorption line surveys targetting circumstellar gas (which would benefit from a new generation UV surveyor).

The CO mass for some systems seems likely high enough that it cannot be explained with a secondary gas scenario; the gas in these systems may be of \hbindex{primordial origin} and may still contain H$_2$ that (together with CO) shield CO from being photodissociated \citep[e.g. in HD 21997,][]{2013ApJ...776...77K}.
In the future it will be important to identify these primordial gas systems.
Understanding why these systems evolved differently from others of similar age will provide vital clues on the transition from the protoplanetary to the debris disc phase (see previous section) and on the origin of debris discs themselves (see previous subsection on the birth of debris discs).
A promising way to identify systems with secondary origin (rather than primordial) is to measure an optically thin CO or CI line ratio (with \hbindex{ALMA} for instance) to check that the gas is out of LTE \citep{2015MNRAS.447.3936M}, i.e. to show that the disc does not contain the abundant H$_2$ colliders expected in a protoplanetary disc \citep{2017MNRAS.464.1415M}.

\subsection{White dwarf polluted discs}

\begin{figure}
\includegraphics[scale=.45]{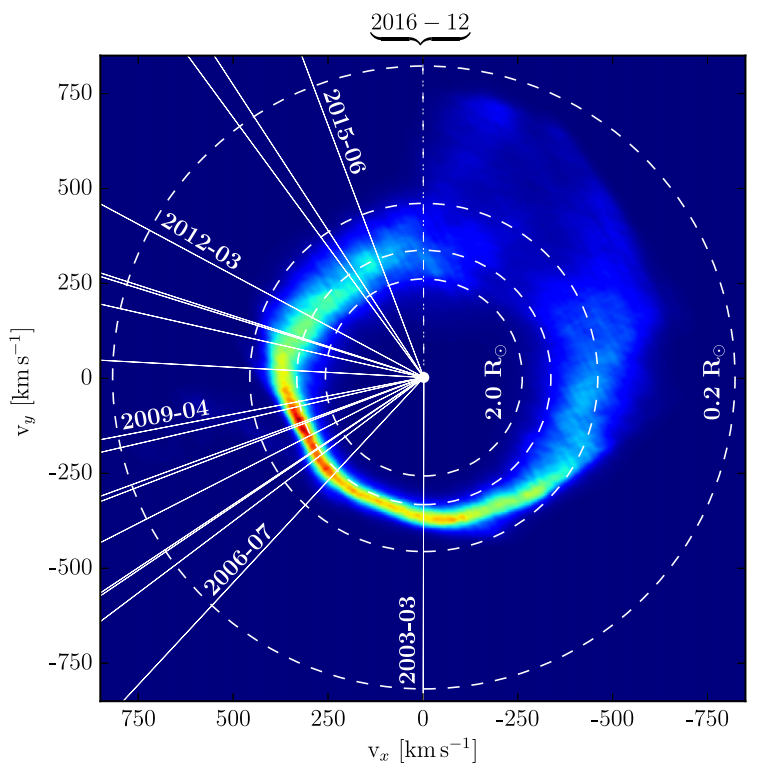}
\caption{An intensity distribution in velocity space of the Ca II triplet which models the line profiles observed in SDSS J1228+1040, obtained from Doppler
tomography \citep{2016MNRAS.455.4467M}.}
\label{fig:wd}       
\end{figure}

Practically all known planet host stars (including our Sun) will evolve into \hbindex{white dwarfs} (WD).
Spectra show that the atmospheres of $\sim$ 30\% of WD are polluted by metals which should not be present due to the short sinking times \citep{2014A&A...566A..34K}.
The best explanation for this pollution is that it comes from the \hbindex{tidal disruption} of planetesimals originally residing in a cold outer belt, which may have been dynamically perturbed by surviving planets \citep{2016NewAR..71....9F,2016RSOS....350571V}.
This is supported by the fact that roughly $2\%$ of WDs also show an IR-excess consistent with circumstellar rings of dust orbiting close to the tidal disruption radius for these stars \citep{2017Bonsor}.
Moreover for one system (WD 1145+017), regular occultations of the star suggest the presence of planetesimals close to the tidal disruption (Roche) limit \citep{2015Natur.526..546V}.
In addition to the absorption lines that are characteristic of WD pollution, gas emission lines are observed for $\sim$10 WDs, inferred to originate in gas that is both very close to the WD and varies with time (see Fig.~\ref{fig:wd}).

Despite a growing body of observational evidence, our understanding of the processes leading to the accretion of planetesimals is poorly understood.
Nevertheless, observations of WD \hbindex{pollution} provide key constraints on the \hbindex{mineralogy} of rocky exo-planetary material.
This is because photospheric metal abundances should trace the bulk composition of accreted planetesimals \citep{2007ApJ...671..872Z}.
For example, by providing key ratios such Mg/Si for the accreted planetesimals, WD pollution measurements provide strong evidence for differentiation in planetary building blocks \citep{2014AREPS..42...45J}, which supports models for planet formation \citep[e.g.][]{2015Icar..247..291B,2015ApJ...813...72C}.
Such measurements also provide a potential opportunity to search for signatures of geological processes (e.g. plate tectonics) in exo-planetary systems, and complement the on-going programs to detect terrestrial planets over the next decade (see subsection on the links between debris and planets).

For now, $\sim$15 WDs have at least five detected pollutant elements in their atmospheres \citep{2014AREPS..42...45J}.
This number is set to triple in the next decade from current and on-going observational programs on HST and VLT (X-shooter), but would benefit greatly from a new FUV mission.
Ground-based observations can reveal Ca (as well as Fe, Mg and Ni) abundances for large samples of WDs.
Such large WD samples are currently being provided by SDSS, but over the next 5 years Gaia will identify $\sim$200,000 WDs brighter than 20th magnitude within 300pc.
Spectroscopic follow-up (e.g., with DESI/4MOST/WEAVE) expects to find 1000's of polluted WD (300 are known today), providing abundances for Ca and/or Mg.
Moreover, after Gaia DR2 (expected April, 2018), we should be able to determine the ages for a large fraction of the $>10^4$ WDs found to a precision of 1-2\%. This will
help to constrain whether the accretion rates, abundance patterns or IR-excesses observed depend on the WD cooling age. 
Finally, after Gaia's final data release and comparative analysis of double WDs and WDs in open clusters, we should be able to derive absolute ages for a large fraction of WDs to $\sim$2\% accuracy \citep{2015ASPC..493..107V}.

Cross correlating these newly detected WDs from Gaia and the \hbindex{AllWISE} catalogue might lead to detections of IR-excesses around these farther WDs.
Follow-up of infrared excesses with JWST/MIRI will also provide direct information on the dust mineralogy that can be compared with abundances found with UV spectra.
Detection of the outer planetesimal belt that is feeding the accretion has been elusive because these belts are expected to be cold \citep{2014MNRAS.444.1821F}.
However, this may be possible with the potential SPICA mission \citep[see Fig. 11 in][]{2010MNRAS.409.1631B}.
In the next 5 years, Gaia will also provide the first real insights into the population of planets around evolved stars \citep{2015ASPC..493..455S} for which there are currently few detections so far \citep{2015A&A...579L...8X}.
Although Gaia can only detect (greater than) Jupiter mass planets, the presence or absence of a correlation between these planets and WD pollution will constrain how WD pollution arises.

The detection of transits blocking the star-light from the polluted white dwarf WD 1145+017 has opened a new window onto the origin of WD pollution and the future detection of similar objects (possible even with relatively modest ground-based telescopes), but presumably in greater abundance with PLATO, will revolutionise our understanding of the fate of planetary systems. 
Continued monitoring of variability in polluted WDs (gas, dust and transits), coupled with 
detailed modelling, will improve our understanding of how planetary material is accreted onto WDs. 

\begin{acknowledgement}
QK and MW acknowledge support from the European Union through ERC grant number 279973. CJC acknoweldges support from the DISCSIM project, grant agreement 341137 funded by the European Research Council under ERC-2013-ADG. QK thanks A. Bonsor for fruitful discussions
about polluted white dwarfs.
\end{acknowledgement}


\end{document}